\documentclass[conference]{IEEEtran}
\IEEEoverridecommandlockouts
\usepackage{cite}
\usepackage{amsmath,amssymb,amsfonts,amsthm}
\usepackage{algorithmic}
\usepackage{graphicx}
\usepackage{textcomp}
\usepackage{accents}
\usepackage{bbm}
\usepackage{mathtools}
\newcommand{\ubar}[1]{\underaccent{\bar}{#1}}
\usepackage{times}
\usepackage{multicol}
\usepackage{}
\def\BibTeX{{\rm B\kern-.05em{\sc i\kern-.025em b}\kern-.08em
    T\kern-.1667em\lower.7ex\hbox{E}\kern-.125emX}}
\newcommand{\euler}{e}
\newcounter{mytempeqncnt}
\begin{document}

\title{Strong Secrecy for General Multiple-Access Wiretap Channels\\
\thanks{This research is co-financed by Greece and the European Union (European Social Fund- ESF) through the Operational Programme ``Human Resources Development, Education and Lifelong Learning'' in the context of the project ``Strengthening Human Resources Research Potential via Doctorate Research'' (MIS-5000432), implemented by the State Scholarships Foundation (IKY).}
}

\author{\IEEEauthorblockN{Manos Athanasakos, and Nicholas Kalouptsidis}
\IEEEauthorblockA{\textit{Department of Informatics and Telecommunications} \\
\textit{National and Kapodistrian University of Athens}\\
Athens, Greece \\
emathan@di.uoa.gr, kalou@di.uoa.gr}
}

\maketitle

\begin{abstract}
This paper is concerned with the general multiple access wiretap channel (MAC-WT) and the existence of codes that accomplish reliability and strong secrecy. Information leakage to the eavesdropper is assessed by the variational distance metric, whereas the average error probability is bounded by modifying Feinstein's Lemma. We derive an achievable strong secrecy rate region by utilizing the resolvability theory for the multiple-access channel in terms of information-spectrum methods.
\end{abstract}


\section{Introduction}


The multiple access wiretap channel involves transmissions from multiple sources to a single destination under the presence of an unauthorized node. Achievable rates are supported by codes that offer vanishingly small error probability at the legitimate receiver and vanishingly small information leakage to the eavesdropper. When the multiple access dimension is separated from the wiretap dimension, well known characterizations of capacity are available for both specialized and abstract channel settings\cite{Cover}. More precisely, the seminal work of Wyner \cite{Wyner}, Csisz\'{a}r and K\"{o}rner \cite{Csi&Kor01} laid the foundations for the analysis of the wiretap channel using information-theoretic tools and proved the existence of coding schemes that achieve reliability and secrecy when rates are confined to the secrecy capacity region. The capacity region of the multiple access channel was extensively studied in \cite{Alshwede}, \cite{Liao}.

During the last decades many communication models have been investigated \cite{Bloch01}, \cite{Poor}, \cite{Yener} using the notion of channel capacity as basis for the development of coding schemes. Recently, the authors in \cite{Bloch&Laneman} showed that capacity based codes have many limitations and cannot provide strong secrecy results. On the other hand, information spectrum methods as laid out in Hayashi \cite{Hayashi} and Han \cite{Han01}, \cite{Han&Verdu} demonstrate that there is a connection between wiretap coding and channel resolvability which can lead to strong secrecy results. The idea behind channel-resolvability based constructions is to use a codebook to control the statistical distribution induced at the output of a noisy channel and the objective is to approximate the eavesdropper's observation by drawing codewords uniformly at random from a codebook. Based on this concept various communication models have been investigated \cite{Bloch&Laneman}, \cite{Yassae}, \cite{Boche} with the strong secrecy requirement as introduced by Maurer \cite{Maurer}.


In this paper, we derive regions of achievable rate pairs for the general multiple-access wiretap channel (MAC-WT) under the strong secrecy criterion. We use the variational distance metric in order to prove statistical independence between the transmitted messages and the eavesdropper distribution, while reliability is provided by modifying Feinstein's Lemma \cite{Feinstein}. Our main result is valid for general channels with abstract alphabets as the proof is based on Steinberg's multiple-access channel (MAC) resolvability theory \cite{Steinberg} and thus can be applied in both a discrete and continuous environment making no common and restrictive assumptions such as stationarity, ergodicity and memory. 

The rest of the paper is organized as follows. Section II describes the model and formulates the problem. In Section III we first introduce some basic prerequisites and then we prove the main theorem highlighting the reliability and information leakage analysis. Finally, Section V concludes the paper.

\textit{Notation:} We represent random variables with capital letters. Sample values are denoted with lower case letters $x$; they take values in alphabets $\mathcal{X}$ which can be arbitrary sets. We use bold upper case letters such as $\mathbf{X}$ to denote random vectors. We write  $p_{\bar{Y}}$  to express the approximated output distribution of a channel resulting from input $\bar{X}$ chosen from code $\mathcal{C}_n$. Also we use the notation $\mathbf{x_1}(i,j)$ and $\mathbf{x}^{(1)}_{ij}$ interchangeably for readability purposes.

\section{Multiple access wiretap channel and coding scheme}

Two users wish to send messages to a legitimate receiver reliably with no errors and simultaneously ensure no leakage to a potential eavesdropper. More precisely, transmission over the MAC-WT, involves two arbitrary input alphabets $\mathcal{X}_1, \mathcal{X}_2$, two arbitrary output alphabets $\mathcal{Y},\mathcal{Z}$ and a sequence of transition probabilities $\{p_{Y Z|X_1 X_2} \}_{n\geq1}$. The channels $(\mathcal{X}_1, \mathcal{X}_2, p(y|x_1,x_2),\mathcal{Y})$ and $(\mathcal{X}_1, \mathcal{X}_2, p(z|x_1,x_2),\mathcal{Z})$ are called the main and the eavesdropper's channel, respectively.\\
 A $(e^{nR_1}, e^{nR_2}, n)$ wiretap coding scheme $\mathcal{C}_n$ for the MAC-WT channel consists of: 
\begin{itemize}
	\item two private message sets $\mathcal{M}_1 = \{1,\ldots,e^{nR_1}\}$ and $\mathcal{M}_2 = \{1,\ldots,e^{nR_2}\}$,
	\item two auxiliary message sets $\mathcal{M}'_1 = \{1,\ldots,e^{nR'_1}\}$ and $\mathcal{M}'_2 = \{1,\ldots,e^{nR'_2}\}$, which are used to randomize the transmission of the private messages,
	\item two stochastic encoders with mappings $\phi_1 : \mathcal{M}_1 \times \mathcal{M}'_1 \to \mathcal{X}_1^n$ and $\phi_2 : \mathcal{M}_2 \times \mathcal{M}'_2 \to \mathcal{X}_2^n,$
	\item and a decoder $\psi_n : \mathcal{Y}^n \to \mathcal{M}_1 \times \mathcal{M}'_1 \times \mathcal{M}_2 \times \mathcal{M}'_2$.\\
\end{itemize}
The reliability performance of the code is evaluated by the probability of error, where we require correct decoding at the legitimate receiver for both the actual and the auxiliary message and is defined as 
\begin{equation*}
\begin{split}
P_e(\mathcal{C}_n) &= P( \{ (\hat{M}_1,\hat{M}_2,\hat{M'}_1,\hat{M'}_2) \\
&\neq (M_1,M_2,M'_1,M'_2) | \mathcal{C}_n \}).
\end{split}
\end{equation*}
Ideally, perfect secrecy obtains when knowledge of the observation vector $\mathbf{Z} \in \mathcal{Z}^n$ at the eavesdropper offers no clues about the transmitted messages $M_1\in \mathcal{M}_1$, $M_2\in \mathcal{M}_2$. Perfect secrecy conditions are asymptotically met provided the relative entropy $D(p_{M_1 M_2 \mathbf{Z}} || p_{M_1 M_2} p_{\mathbf{Z}})$ goes to zero as $n\to \infty$. Here we will assess the amount of leaked information to the eavesdropper by the second strongest secrecy metric \cite{Bloch&Laneman} defined in terms of the variational distance.
\begin{equation*}
\mathcal{L}(\mathcal{C}_n) = V(p_{M_1 M_2 \mathbf{Z}}, p_{M_1 M_2} p_{\mathbf{Z}}).
\end{equation*}
 Where $V$ is the variational distance between two distributions\\
\begin{equation*}
V(P,Q) = 2\sup_{\mathcal{A}\in\mathcal{F}} |P(\mathcal{A}) - Q(\mathcal{A})|
\end{equation*}
Based on the above, we call a pair $(R_1,R_2)$ achievable if there exists a sequence of codes such that the 
\begin{itemize}
	\item $\lim_{n \to \infty} P_e(\mathcal{C}_n) = 0, $
	\item  $\lim_{n \to \infty} \mathcal{L}(\mathcal{C}_n) = 0$.
\end{itemize}
The strongly secure achievable rate region of the general MAC-WT channel is characterized in the next section.
\section{Achievability of the MAC-WT channel}
The achievable rate region of the MAC-WT channel is described in terms of information density and spectrum concepts. These are defined next.
\subsection{Information-spectrum theory}
The information density between $X$ and $Y$ is the random variable
\begin{equation*}
i(X;Y) = \log \frac{p_{XY}(XY)}{p_{X}(X)p_{Y}(Y)}
\end{equation*}
and the conditional information density of $X$, $Y$ given $Z$ is defined as
\begin{equation*}
i(X;Y|Z) = \log \frac{p_{XY|Z}(X,Y|Z)}{p_{X|Z}(X|Z)p_{Y|Z}(Y|Z)}.
\end{equation*}
The mean of the above variables is the usual mutual information, i.e. $I(X;Y) = E_{XY}[i(X;Y)]$ and $I(X;Y|Z) = E_{XYZ}[i(X;Y|Z)]$.
If $\mathbf{X} = (X_1,\ldots,X_n)$ and $\mathbf{Y} = (Y_1,\ldots,Y_n)$ denote random vectors of dimension $n$, the distribution of the random variable $\frac{1}{n}i(\mathbf{X};\mathbf{Y})$ is called the \textit{information spectrum} \cite{Han01} of $p_{\mathbf{X}\mathbf{Y}}$. The asymptotic regime is described by the \textit{sup-information rate} 
\begin{equation*}
\begin{split}
\bar{i}(\mathbf{X;Y}) &= \text{p -}\limsup_{n\to \infty} \frac{1}{n} i(\mathbf{X;Y}) \\
&= \inf \{\alpha: \lim_{n\to \infty}P[\frac{1}{n} i(\mathbf{X;Y}) > \alpha] = 0  \},
\end{split}
\end{equation*}
and the \textit{inf-information rate} 
\begin{equation*}
\begin{split}
\ubar{i}(\mathbf{X;Y}) &= \text{p -}\liminf_{n\to \infty} \frac{1}{n} i(\mathbf{X;Y}) \\
&= \sup \{\beta: \lim_{n\to \infty}P[\frac{1}{n} i(\mathbf{X;Y}) < \beta] = 0  \}.
\end{split}
\end{equation*}
 We are now ready to present the main result of this paper.
\textbf{Theorem:} The achievable rate region under the strong secrecy criterion for the general MAC-WT is
\begin{equation*}
\mathcal{R}_{SS}=
\begin{cases}
	\!\begin{aligned}
		R_1 \leq \text{p-}&\liminf_{n\to \infty} \frac{1}{n} i(\mathbf{X_1;Y|X_2})\\ &-\text{p-}\limsup_{n\to \infty} \frac{1}{n} i(\mathbf{X_1 ;Z})- 4\gamma\\
		R_2 \leq \text{p-}&\liminf_{n\to \infty} \frac{1}{n} i(\mathbf{X_2;Y|X_1})\\ &-\text{p-}\limsup_{n\to \infty} \frac{1}{n} i(\mathbf{ X_2;Z}) - 4\gamma\\
		R_1+R_2 \leq \text{p-}&\liminf_{n\to \infty} \frac{1}{n} i(\mathbf{X_1 X_2;Y})\\ &-\text{p-}\limsup_{n\to \infty} \frac{1}{n} i(\mathbf{X_1 X_2;Z}) - 4\gamma
	\end{aligned}
\end{cases}
\end{equation*}
where $\gamma$ is a positive arbitrary constant.\\
\textit{Remark:} In \cite{Yassae} Yassaee and Aref, use output statistics approximation techniques  to obtain the same rate region for the MAC-WT. However, their result is confined to discrete memoryless channels as it is based on typicality. In this paper achievability is established for a broader class of channels that includes both discrete and continuous channel and without memory constraints.\\
\newline
The proof of the above theorem proceeds in a number of steps. First we analyze achievability by working out the attributes that shape the probability of error. Then we demonstrate that in addition the strong secrecy criterion is valid. More precisely, consider a rate pair $(R_1,R_2)$ satisfying the inequalities given in the theorem. We show that there is a sequence of codes with asymptotically vanishing error probability and leakage.
\subsection{Reliability analysis}
We construct  a coding scheme with the following structure:\\
\textit{Codebook generation:} We generate  $M_1 M'_1$ codewords independently and at random by generating $\lceil e^{nR_1}\rceil \lceil e^{nR'_1} \rceil$ i.i.d. sequences $x^n_1(m_1)$ with $m_1 = (i,j) \in \mathcal{M}_1 \times \mathcal{M}'_1$ according to $p_{X_1}$. Similarly we generate the codewords for the second user.\\
\textit{Encoding:} To transmit $m_1 = (i,j) \in \mathcal{M}_1 \times \mathcal{M}'_1$, where $i$ is the actual message and $j$ is the auxiliary message, user 1 computes $x^n_1(m_1)$ and send it over the channel. User 2 acts in the same way.\\
\textit{Decoding:} For arbitrary $\gamma>0$ we define the sets 
\begin{equation*}
\begin{split}
\mathcal{T}_1^n &= \{(\mathbf{x_1},\mathbf{x_2},\mathbf{y}):\\
&\frac{1}{n} \log \frac{p_{\mathbf{Y}|\mathbf{X_1} \mathbf{X_2}}(\mathbf{y}|\mathbf{x_1},\mathbf{x_2})}{p_{\mathbf{Y}|\mathbf{X_2}}(\mathbf{y}|\mathbf{x_2})} > \frac{1}{n} \log M_1 M'_1 + \gamma \bigg\},\\
\mathcal{T}_2^n &= \{(\mathbf{x_1},\mathbf{x_2},\mathbf{y}):\\
&\frac{1}{n} \log \frac{p_{\mathbf{Y}|\mathbf{X_1} \mathbf{X_2}}(\mathbf{y}|\mathbf{x_1},\mathbf{x_2})}{p_{\mathbf{Y}|\mathbf{X_1}}(\mathbf{y}|\mathbf{x_1})} > \frac{1}{n} \log M_2 M'_2 + \gamma \bigg\},\\
\mathcal{T}_{12}^n &= \{(\mathbf{x_1},\mathbf{x_2},\mathbf{y}):\\
&\frac{1}{n} \log \frac{p_{\mathbf{Y}|\mathbf{X_1} \mathbf{X_2}}(\mathbf{y}|\mathbf{x_1},\mathbf{x_2})}{p_{\mathbf{Y}}(\mathbf{y})} > \frac{1}{n} (\log M_1 M_2 + \log M'_1 M'_2)\\
&+ \gamma \bigg\}
\end{split}
\end{equation*}
and  $\mathcal{T}^n = \mathcal{T}_1^n \cap \mathcal{T}_2^n \cap \mathcal{T}_{12}^n$.\\
 After receiving a $\mathbf{y}\in \mathcal{Y}^n$, the threshold decoder $\psi_n$ searches for messages $m_1, m_2$ and their corresponding indices which satisfy $(\mathbf{x_1}(m_1), \mathbf{x_2}(m_2), y^n)\in \mathcal{T}^n$. If such a tuple exists and is unique the decoder outputs $\psi_n(\mathbf{y}) = (\hat{m}_1, \hat{m}_2)$, otherwise declares an error.\\

The analysis of the error probability relies on standard arguments. The following Lemma provides conditions for reliable communication between the two users and the legitimate receiver.\\
\textbf{Lemma 1:} If  
\begin{eqnarray*}
\begin{split}
0 \leq R_1 + R'_1 &\leq \text{p-}\liminf_{n\to \infty} \frac{1}{n} i(\mathbf{X_1;Y|X_2}) - 2\gamma\\
0 \leq R_2 + R'_2 &\leq \text{p-}\liminf_{n\to \infty} \frac{1}{n} i(\mathbf{X_2;Y|X_1}) - 2\gamma\\
0 \leq R_1 +  R_2 + R'_1 + R'_2 &\leq \text{p-}\liminf_{n\to \infty} \frac{1}{n} i(\mathbf{X_1 X_2;Y}) - 2\gamma,
\end{split}
\end{eqnarray*}
then $\lim_{n \to \infty} E[P_e(\mathcal{C}_n)] = 0$.
\begin{proof}
In order to prove that for the any rate pairs which comply with the above inequalities, $P_e(\mathcal{C}_n)\rightarrow 0$ for large $n$, we consider the multi-user version of Feinstein's Lemma \cite{Feinstein}, \cite{Han01} and modify it properly so that includes the auxiliary message rates. This is necessary since the legitimate receiver has to decode the auxiliary random message as well. More precisely the following Lemma holds.\\
\textbf{Lemma 2:} Let $X_1$, $X_2$ be an arbitrary pair of channel inputs and $Y$ is the channel output from a MAC. Then for arbitrary positive integers $M_1,M_2, M'_1,M'_2$, there exists an $(M_1,M_2,M'_1,M'_2,n)$-code satisfying
\begin{eqnarray*}
\begin{split}
P_e(\mathcal{C}_n) & \leq P\bigg\{ \frac{1}{n} i(\mathbf{X_1};\mathbf{Y}|\mathbf{X_2}) \leq \frac{1}{n} \log M_1 M'_1 + \gamma \bigg\} \\
&+ P\bigg\{ \frac{1}{n} i(\mathbf{X_2};\mathbf{Y}|\mathbf{X_1}) \leq \frac{1}{n} \log M_2 M'_2 + \gamma  \bigg\}\\
&+ P\bigg\{ \frac{1}{n} i(\mathbf{X_1}\mathbf{X_2};\mathbf{Y}) \leq \frac{1}{n}  (\log M_1 M_2 + \log M'_1 M'_2)\\
&+ \gamma  \bigg\}+ 5e^{-n\gamma}.
\end{split}
\end{eqnarray*}
for all $n$ and arbitrary $\gamma>0$.\\
We provide a sketch of the proof only; the full details follow the same line of arguments with those employed for the simple MAC case in \cite[Theorem 7.7]{Han01}. First we define the event
\begin{equation*}
\mathcal{E}_{ijkl} = {(\mathbf{x_1}(i,j),\mathbf{x_2}(k,l),\mathbf{y})\in\mathcal{T}^n}
\end{equation*}
 $\mathcal{E}^c_{ijkl}$ stands for the complement. Using the symmetry of the random code construction, the average error probability $E[P_e(\mathcal{C}_n)]$ is equal to the error probability of a random codeword transmission. Without loss of generality, we assume that user 1 and user 2 send $\mathbf{x_1}(1,1)$ and $\mathbf{x_2}(1,1)$, respectively. Therefore the error probability is upper bounded as follows:
\begin{eqnarray}\label{eq2}
\begin{split}
&E[P_e(\mathcal{C}_n)] = P\{\mathcal{E}^c_{1111} \bigcup\limits_{\substack{(i'j'k'l')\\ \neq \\ (1111)}} \mathcal{E}_{i'j'k'l'} \}\\
&\leq  P\{\mathcal{E}^c_{1111}\} + \sum_{i'\neq 1} P\{\mathcal{E}_{i'111}\} + \sum_{j'\neq 1} P\{\mathcal{E}_{1j'11}\}\\
&+\sum_{k'\neq 1} P\{\mathcal{E}_{11k'1}\} + \sum_{l'\neq 1} P\{\mathcal{E}_{111l'}\} + \sum_{\substack{(i'j'k'l')\\ \neq \\ (1111)}} P\{\mathcal{E}_{i'j'k'l'} \} 
\end{split}
\end{eqnarray}
Each of the terms appearing above can be handled in a manner similar to \cite[Theorem 7.7]{Han01}. For instance, the first summand is bounded as follows
\begin{eqnarray}\label{eq3}
\begin{split}
P\{\mathcal{E}_{i'111}\} &= \sum_{(\mathbf{x_1},\mathbf{x_2},\mathbf{y})\in \mathcal{T}^n}p_{\mathbf{X_1}}(\mathbf{x_1}) p_{\mathbf{X_2 Y}}(\mathbf{x_2},\mathbf{y})\\
&\leq \sum_{(\mathbf{x_1},\mathbf{x_2},\mathbf{y})\in \mathcal{T}_1^n}p_{\mathbf{X_1}}(\mathbf{x_1}) p_{\mathbf{X_2 Y}}(\mathbf{x_2},\mathbf{y}).
\end{split}
\end{eqnarray}
Also, for $(\mathbf{x_1},\mathbf{x_2},\mathbf{y})\in \mathcal{T}_1^n$
\begin{equation}\label{eq4}
p_{\mathbf{X_2 Y}}(\mathbf{x_2},\mathbf{y})\leq p_{\mathbf{X_2}}(\mathbf{x_2}) p_{\mathbf{Y}|\mathbf{X_1}\mathbf{X_2}}(\mathbf{y}|\mathbf{x_1}\mathbf{x_2}) \frac{e^{-n\gamma}}{M_1}
\end{equation}
and eq.(\ref{eq2}) leads to 
\begin{equation}
\begin{split}
P\{&\mathcal{E}_{i'111}\}\\  
&\leq \sum_{(\mathbf{x_1},\mathbf{x_2},\mathbf{y})\in \mathcal{T}_1^n} p_{\mathbf{X_1}}(\mathbf{x_1}) p_{\mathbf{X_2}}(\mathbf{x_2}) p_{\mathbf{Y}|\mathbf{X_1}\mathbf{X_2}}(\mathbf{y}|\mathbf{x_1}\mathbf{x_2}) \frac{e^{-n\gamma}}{M_1}\\
&\leq \frac{e^{-n\gamma}}{M_1},
\end{split}
\end{equation}
and finally we obtain $\sum_{i'\neq 1}P\{\mathcal{E}_{i'111}\} \leq e^{-n\gamma}$. Similarly we can bound the other terms and by substituting into eq. (\ref{eq2}) we get 
\begin{equation*}
E[P_e(\mathcal{C}_n)] \leq P\{\mathcal{E}^c_{1111}\} + 5e^{-n\gamma}
\end{equation*}
where the first term is trivial.
According to this we conclude that there must exist at least one $(M_1,M_2,M'_1,M'_2,n)$-code satisfying the condition of Lemma 2 and any information and auxiliary rate pair satisfying Lemma 1 is achievable.
Hence, for arbitrary $(R_1,R_2)$ and $(R'_1,R'_2)$ and a small constant $\gamma > 0$ we
define $M_i = e^{n(R_i-2\gamma)}$ and $M'_i = e^{n(R'_i-2\gamma)}$ for $i=1,2$ and Lemma 2 guarantees the existence of an $(M_1,M_2,M'_1,M'_2,n)$-code satisfying
\begin{eqnarray*}
\begin{split}
P_e(\mathcal{C}_n) & \leq P\bigg\{ \frac{1}{n} i(\mathbf{X_1};\mathbf{Y}|\mathbf{X_2}) \leq  R_1 + R'_1 - 3\gamma \bigg\} \\
&+ P\bigg\{ \frac{1}{n} i(\mathbf{X_2};\mathbf{Y}|\mathbf{X_1}) \leq  R_2 + R'_2 - 3\gamma  \bigg\}\\
&+ P\bigg\{ \frac{1}{n} i(\mathbf{X_1}\mathbf{X_2};\mathbf{Y}) \leq  R_1 + R'_1 + R_2 + R'_2 \\
&- 7\gamma  \bigg\} + 5e^{-n\gamma}.
\end{split}
\end{eqnarray*}
Because the definitions of spectral inf-information rates imply that all the terms of the RHS converge to zero as $n\rightarrow \infty$ then $\lim\limits_{n\to \infty}E[P_e(\mathcal{C}_n)]=0$.
\end{proof}

\subsection{Secrecy Analysis}
We next analyze the code performance in terms of information leakage. The following Lemma identifies the constraints on the rates of local randomness needed to asymptotically eliminate leakage. \\
\textbf{Lemma 3: } If the rate pair satisfies
\begin{equation}
\begin{split}
R'_1 &\geq \text{p-}\limsup_{n\to \infty} \frac{1}{n} i(\mathbf{X_1 ;Z}) + 2\gamma\\
R'_2 &\geq \text{p-}\limsup_{n\to \infty} \frac{1}{n} i(\mathbf{ X_2;Z}) + 2\gamma\\
R'_1+R'_2 &\geq \text{p-}\limsup_{n\to \infty} \frac{1}{n} i(\mathbf{X_1 X_2;Z}) + 2\gamma
\end{split}
\end{equation}
then $\lim_{n \to \infty} E[\mathcal{L}(\mathcal{C}_n)] = 0$.
\begin{proof}
Let $\bar{\mathbf{Z}}$ denote the random vector received by the eavesdropper if $\mathbf{X_1}$, $\mathbf{X_2}$ are i.i.d distributed according to $p_{\mathbf{X_1}}(\mathbf{x_1})$, $p_{\mathbf{X_2}}(\mathbf{x_2})$. We can bound the variational distance between $p_{M_1 M_2 \mathbf{Z}}$ and $p_{M_1 M_2} p_{\mathbf{Z}}$ as follows
\begin{equation}
\begin{split}
	\mathcal{L}(\mathcal{C}_n) &= V(p_{M_1 M_2 \bar{\mathbf{Z}}},p_{M_1 M_2} p_{\bar{\mathbf{Z}}}) \\
	&= E_{M_1 M_2}[V(p_{\bar{\mathbf{Z}}|M_1 M_2},p_{\bar{\mathbf{Z}}} )] \\
	& \leq E_{M_1 M_2}[ V(p_{\bar{\mathbf{Z}}|M_1 M_2}, p_{\mathbf{Z}}) + V(p_{\mathbf{Z}},p_{\bar{\mathbf{Z}}})] \\
	& \leq 2E_{M_1 M_2}[V(p_{\bar{\mathbf{Z}}|M_1 M_2}, p_{\mathbf{Z}})].
\end{split}
\end{equation}
Furthermore if we average over all possible codewords and use the symmetry of the random coding construction, we obtain
\begin{equation}\label{eq4}
E[\mathcal{L}(\mathcal{C}_n)] \leq 2E[V(p_{\bar{\mathbf{Z}}|M_1=1 M_2=1}, p_{\mathbf{Z}})]
\end{equation}
where the output distribution is given by
\begin{equation}
p_{\bar{\mathbf{Z}}|11}(\mathbf{z}) = \frac{1}{M'_1 M'_2} \sum_{j=1}^{M'_1} \sum_{l=1}^{M'_2} p_{\mathbf{Z}|\mathbf{X_1} \mathbf{X_2}}(\mathbf{z}|\mathbf{x_1}(1,j),\mathbf{x_2}(1,l))
\end{equation}
To ensure asymptotic decay of $\mathcal{L}(\mathcal{C}_n)$ to zero, it suffices to establish that each of the sub-codebooks $\{\mathbf{x_1}(i,j)\}_{j\in M'_1}$ and $\{\mathbf{x_2}(k,l)\}_{l\in M'_2}$ are channel resolvability codes \cite{Bloch01},\cite{Han01}. The proof of this claim relies on the following known inequality.\\
\textbf{Lemma 4:}\cite{Han01} For any $\mu>0$ and any two distributions $P$,$Q$
\begin{equation*}
V(P,Q) \leq 2\mu + 2  \underbrace{P\bigg\{ \log \frac{P(X)}{Q(X)} > \mu \bigg\}}_{\mathcal{J}_\mu}
\end{equation*}
where $P$, $Q$ are distributed over $X$.\\
\begin{figure*}[!t]
	\normalsize
	\setcounter{mytempeqncnt}{\value{equation}}
	\setcounter{equation}{10}
	\begin{equation} \label{eq00}
	\begin{split}
	 &\mathcal{J}_\mu= \sum_{x^{(1)}_{1 2} \in \mathcal{X}^{n}_1} \ldots \sum_{x^{(1)}_{1 M'_1} \in \mathcal{X}^{n}_1} \sum_{x^{(2)}_{1 2} \in \mathcal{X}^{n}_2} \ldots \sum_{x^{(2)}_{1 M'_2} \in \mathcal{X}^{n}_2} p_{\mathbf{X_1}}(\mathbf{x}^{(1)}_{1 2}) \ldots
	p_{\mathbf{X_1}}(\mathbf{x}^{(1)}_{1 M'_1}) p_{\mathbf{X_2}}(\mathbf{x}^{(2)}_{1 2}) \ldots p_{\mathbf{X_2}}(\mathbf{x}^{(2)}_{1 M'_2}) \\
	&\times \sum_{x^{(1)}_{1 1} \in \mathcal{X}^{n}_1} \sum_{x^{(2)}_{1 1} \in \mathcal{X}^{n}_2} \sum_{z\in \mathcal{Z}^n} p_{\mathbf{X_1}\mathbf{X_2}\mathbf{Z}}(\mathbf{x}^{(1)}_{1 1},\mathbf{x}^{(2)}_{1 1},\mathbf{z}) \mathbbm{1} \bigg\{ \frac{1}{M'_1 M'_2} \exp(i(\mathbf{x}^{(1)}_{1 1} \mathbf{x}^{(2)}_{1 1};\mathbf{z}))
	+ \frac{1}{M'_1 M'_2} \sum_{(j,l)\neq (1,1)} \exp(i(\mathbf{x}^{(1)}_{1 j} \mathbf{x}^{(2)}_{1 l};\mathbf{z})) > 1+4\tau \bigg\}\\
	&\leq \sum_{x^{(1)}_{1 2} \in \mathcal{X}^{n}_1} \ldots \sum_{x^{(1)}_{1 M'_1} \in \mathcal{X}^{n}_1} \sum_{x^{(2)}_{1 2} \in \mathcal{X}^{n}_2} \ldots \sum_{x^{(2)}_{1 M'_2} \in \mathcal{X}^{n}_2} p_{\mathbf{X_1}}(\mathbf{x}^{(1)}_{1 2}) \ldots
	p_{\mathbf{X_1}}(\mathbf{x}^{(1)}_{1 M'_1}) p_{\mathbf{X_2}}(\mathbf{x}^{(2)}_{1 2}) \ldots p_{\mathbf{X_2}}(\mathbf{x}^{(2)}_{1 M'_2}) \\
	&\times \sum_{x^{(1)}_{1 1} \in \mathcal{X}^{n}_1} \sum_{x^{(2)}_{1 1} \in \mathcal{X}^{n}_2} \sum_{z\in \mathcal{Z}^n} p_{\mathbf{X_1}\mathbf{X_2}\mathbf{Z}}(x^{(1)}_{1 1},x^{(2)}_{1 1},\mathbf{z})\Bigg\{ \mathbbm{1} \bigg\{ \frac{1}{M'_1 M'_2} \exp(i(\mathbf{x}^{(1)}_{1 1} \mathbf{x}^{(2)}_{1 1};\mathbf{z})) > \tau \bigg\} + \mathbbm{1}\bigg\{ \frac{1}{M'_1 M'_2} \sum_{j\neq 1} \exp(i(\mathbf{x}^{(1)}_{1 j} \mathbf{x}^{(2)}_{1 1};\mathbf{z})) > \tau \bigg\}\\
	&+\mathbbm{1}\bigg\{ \frac{1}{M'_1 M'_2} \sum_{l\neq 1} \exp(i(\mathbf{x}^{(1)}_{1 1} \mathbf{x}^{(2)}_{1 l};\mathbf{z})) > \tau \bigg\} + \mathbbm{1}\bigg\{ \frac{1}{M'_1 M'_2} \sum_{(j,l)\neq (1,1)} \exp(i(\mathbf{x}^{(1)}_{1 j} \mathbf{x}^{(2)}_{1 l};\mathbf{z})) > 1+\tau \bigg\}  \Bigg\}
	\end{split}
	\end{equation}
	\setcounter{equation}{\value{mytempeqncnt}}
	\hrulefill
\end{figure*}
Guided by the above Lemma we define $\mathcal{J}_\mu $ for any $\mu>0$ as follows,
\begin{equation}\label{eq6}
\begin{split}
\mathcal{J}_\mu = &\sum_{x^{(1)}_{1 1} \in \mathcal{X}^{n}_1} \ldots \sum_{x^{(1)}_{1 M'_1} \in \mathcal{X}^{n}_1} \sum_{x^{(2)}_{1 1} \in \mathcal{X}^{n}_2} \ldots \sum_{x^{(2)}_{1 M'_2} \in \mathcal{X}^{n}_2} p_{\mathbf{X_1}}(\mathbf{x}^{(1)}_{1 1}) \ldots \\
&p_{\mathbf{X_1}}(\mathbf{x}^{(1)}_{1 M'_1}) p_{\mathbf{X_2}}(\mathbf{x}^{(2)}_{1 1}) \ldots p_{\mathbf{X_2}}(\mathbf{x}^{(2)}_{1 M'_2}) \\
&\times \sum_{z\in \mathcal{Z}^n} p_{\bar{\mathbf{Z}}|11}(\mathbf{z}) \mathbbm{1} \bigg\{\log \frac{p_{\bar{\mathbf{Z}}|11}(\mathbf{z})}{p_{\mathbf{Z}}(\mathbf{z})} > \mu \bigg\}
\end{split}
\end{equation}
and  we show that for any $\mu>0$; $\mathcal{J}_\mu \rightarrow 0$ \cite{Steinberg}. Indeed,
\begin{equation} \label{eq11}
\begin{split}
\mathcal{J}_\mu = &\sum_{x^{(1)}_{1 1} \in \mathcal{X}^{n}_1} \ldots \sum_{x^{(1)}_{1 M'_1} \in \mathcal{X}^{n}_1} \sum_{x^{(2)}_{1 1} \in \mathcal{X}^{n}_2} \ldots \sum_{x^{(2)}_{1 M'_2} \in \mathcal{X}^{n}_2} p_{\mathbf{X_1}}(\mathbf{x}^{(1)}_{1 1}) \ldots \\
&p_{\mathbf{X_1}}(\mathbf{x}^{(1)}_{1 M'_1}) p_{\mathbf{X_2}}(\mathbf{x}^{(2)}_{1 1}) \ldots p_{\mathbf{X_2}}(\mathbf{x}^{(2)}_{1 M'_2}) \\
&\times \sum_{z\in \mathcal{Z}^n} \sum_{j=1}^{M'_1} \sum_{l=1}^{M'_2} \frac{1}{M'_1 M'_2} p_{\mathbf{Z}|\mathbf{X_1} \mathbf{X_2}}(\mathbf{z}|\mathbf{x_1}(1,j),\mathbf{x_2}(1,l))\\
&\times \mathbbm{1} \bigg\{\frac{\sum_{j=1}^{M'_1} \sum_{l=1}^{M'_2}p_{\mathbf{Z}|\mathbf{X_1} \mathbf{X_2}}(\mathbf{z}|\mathbf{x_1}(1,j),\mathbf{x_2}(1,l))}{M'_1 M'_2 p_{\mathbf{Z}}(\mathbf{z})} > \euler^\mu \bigg\}\\
&= \sum_{x^{(1)}_{1 2} \in \mathcal{X}^{n}_1} \ldots \sum_{x^{(1)}_{1 M'_1} \in \mathcal{X}^{n}_1} \sum_{x^{(2)}_{1 2} \in \mathcal{X}^{n}_2} \ldots \sum_{x^{(2)}_{1 M'_2} \in \mathcal{X}^{n}_2} p_{\mathbf{X_1}}(\mathbf{x}^{(1)}_{1 2}) \ldots \\
&p_{\mathbf{X_1}}(\mathbf{x}^{(1)}_{1 M'_1}) p_{\mathbf{X_2}}(\mathbf{x}^{(2)}_{1 2}) \ldots p_{\mathbf{X_2}}(\mathbf{x}^{(2)}_{1 M'_2}) \\
&\times \sum_{x^{(1)}_{1 1} \in \mathcal{X}^{n}_1} \sum_{x^{(2)}_{1 1} \in \mathcal{X}^{n}_2} \sum_{z\in \mathcal{Z}^n} p_{\mathbf{X_1}\mathbf{X_2}\mathbf{Z}}(\mathbf{x}^{(1)}_{1 1},\mathbf{x}^{(2)}_{1 1},\mathbf{z}) \\
&\times \mathbbm{1} \bigg\{ \frac{p_{\mathbf{Z}|\mathbf{X_1} \mathbf{X_2}}(\mathbf{z}|\mathbf{x_1}(1,1),\mathbf{x_2}(1,1))}{M'_1 M'_2 p_{\mathbf{Z}}(\mathbf{z})}\\
&+ \frac{\sum_{(j,l)\neq (1,1)} p_{\mathbf{Z}|\mathbf{X_1} \mathbf{X_2}}(\mathbf{z}|\mathbf{x_1}(1,j),\mathbf{x_2}(1,l))}{M'_1 M'_2 p_{\mathbf{Z}}(\mathbf{z})} > \euler^\mu \bigg\}
\end{split}
\end{equation}
where the last equality exploits symmetry. Moreover, from eq.(\ref{eq11}) we obtain eq. (11) on the top of the next page, where $\tau = \frac{1}{4}(\euler^{\mu} - 1)$. Taking the expectation we obtain 
\setcounter{equation}{11}
\begin{equation} \label{eq13}
\mathcal{J}_\mu \leq \mathcal{J}_1+\mathcal{J}_2+\mathcal{J}_3+\mathcal{J}_4
\end{equation}
where
\begin{equation*} \label{eq14}
\begin{split}
\mathcal{J}_1 &= P\bigg\{ \frac{1}{M'_1 M'_2}\exp(i(\mathbf{X_1} \mathbf{X_2};\mathbf{Z}))> \tau \bigg\}\\
\mathcal{J}_2 &= P\bigg\{\frac{1}{M'_1 M'_2}\sum_{j=1}^{M'_1}\exp(i(\mathbf{X_1}(1,j) \mathbf{X_2};\mathbf{Z})) > \tau \bigg\}\\
\mathcal{J}_3 &= P\bigg\{\frac{1}{M'_1 M'_2}\sum_{l=1}^{M'_2}\exp(i(\mathbf{X_1} \mathbf{X_2}(1,l);\mathbf{Z})) > \tau \bigg\} \\
\mathcal{J}_4 &= P\bigg\{\frac{1}{M'_1 M'_2}\sum_{j=1}^{M'_1} \sum_{l=1}^{M'_2} \exp(i(\mathbf{X_1}(1,j) \mathbf{X_2}(1,l);\mathbf{Z})) \\
&> 1 + \tau \bigg\},
\end{split}
\end{equation*}
Application of similar steps to those developed in \cite{Steinberg} allows us to conclude that each one the terms $\mathcal{J}_i, i = 1,2,3,4$, go to zero, if the auxiliary rate pairs satisfy the inequalities of Lemma 3. Finally, convergence of $\mathcal{J}_{\mu}$ to zero in conjunction with Lemma 4 prove the claim.
\end{proof}

Finally, combining the results of Lemma 1 and Lemma 3 and performing Fourier-Motzkin elimination for the auxiliary rates $R'_1, R'_2$ we obtain the achievable rate region of the main Theorem.
 
\section{Conclusion}
In this paper achievability has been studied for strongly secure communication in a MAC-WT. Thanks to the information-spectrum methods adopted in this work, there is no need for assumptions such as stationarity, ergodicity and memorylessness. Moreover we obtained generalized results which can be useful to specific applications. Extensions that include relay nodes as well as specialized settings are considered in forthcoming work.


\end{document}